\documentclass[epj]{svjour}
\begin{document}
\title{Probability distribution function for systems driven by
superheavy-tailed noise}
\author{S.I. Denisov\inst{1,2,}\thanks{\rm{e-mail: stdenis@pks.mpg.de}}
\and H. Kantz\inst{1}}
\institute{Max-Planck-Institut f\"{u}r Physik komplexer Systeme, N\"{o}thnitzer
Stra{\ss}e 38, D-01187 Dresden, Germany \and Sumy State University,
Rimsky-Korsakov Street 2, UA-40007 Sumy, Ukraine }
\date{Received:  }
\abstract{We develop a general approach for studying the cumulative probability
distribution function of localized objects (particles) whose dynamics is
governed by the first-order Langevin equation driven by superheavy-tailed
noise. Solving the corresponding Fokker-Planck equation, we show that due to
this noise the distribution function can be divided into two different parts
describing the surviving and absorbing states of particles. These states and
the role of superheavy-tailed noise are studied in detail using the theory of
slowly varying functions.
}

\maketitle

\section{Introduction}
\label{Intr}

The Langevin approach, i.e., incorporation of the noise terms into
deterministic equations of motion to describe the stochastic dynamics of
particles, is one of the most effective tools for studying the effects induced
by fluctuating environment. If these terms arise from the time derivative of
the noise generating processes, i.e., random processes with independent and
identically distributed increments, then the solution $x(t)$ of the
corresponding Langevin equation is a Markov process \cite{HL,CKW}. The main
feature of Markov processes is that their future behavior depends only on the
present state, or in other words, the correlation time of Markov processes
equals zero. Although physical processes are usually characterized by nonzero
correlation times, the Markov processes can be considered as a first
approximation of physical ones if the correlation time is the smallest time
scale of the system.

In the Markovian case, the statistical properties of the particle position
$x(t)$ are controlled by the distribution of increments of the noise generating
process. If this distribution is described by a scaled probability density
$p(y)$ (see the next section) then, depending on the asymptotic behavior of the
tails of $p(y)$, three qualitatively different cases can be distinguished. The
Gaussian density represents the first case in which the process $x(t)$ is
continuous and the probability density $P(x,t)$ of $x(t)$ satisfies the
ordinary Fokker-Planck equation \cite{Risk,Gard}. This equation describes a
wide variety of noise phenomena, including particle diffusion in external
potentials \cite{Risk,Gard}, stochastic resonance \cite{GHJM}, noise-induced
transport \cite{JAP,Reim,HM}, noise-enhanced stability \cite{AS,DAS},
cross-correlation effects \cite{DVH,DVHH}, and many others. The L\'{e}vy stable
densities represent the second case in which the probability density $P(x,t)$
satisfies the fractional Fokker-Planck equation \cite{JMF,Dit,SLDYL,BS,DS,DHH1}
that describes a special class of discontinuous processes $x(t)$, the so-called
L\'{e}vy flights \cite{SZF,MK,CGKM,DSU}. It has been recently shown \cite{DHK}
that the ordinary Fokker-Planck equation holds for all probability densities
$p(y)$ with finite variance, and the fractional one for all heavy-tailed
$p(y)$, i.e., probability densities with power-law tails and infinite second
moment.

Finally, the third case corresponds to the superheavy-tailed $p(y)$, i.e.,
probability densities whose all fractional moments are infinite. Because of
this unusual feature, the distributions with superheavy tails are rarely used
in phy\-sics. Nevertheless, the examples of ultraslow diffusion
\cite{HW1,HW2,DK,DeKa} demonstrate the utility of these distributions in
modeling the systems with an extremely anomalous behavior. Very recently, the
usefulness of superheavy-tailed distributions has also been demonstrated for
the Langevin approach \cite{DKH}. Specifically, using the generalized
Fokker-Planck equation \cite{DHH1}, it has been shown that superheavy-tailed
noise, i.e., noise arising from a generating process whose independent
increments are distributed with superheavy tails, induces two probabilistic
states of a particle, surviving and absorbing. Due to this feature, the
Langevin equation driven by superheavy-tailed noise can be used to describe
some randomly interrupted processes. From a physical point of view, the
interruption can be associated with the transition of a particle to a
qualitatively new state. In particular, such a situation occurs when the
particle dynamics is accompanied by the absorption of these particles. It
should be stressed, however, that the Langevin equation driven by
superheavy-tailed noise and the corresponding generalized Fokker-Planck
equation describe the simplest situation when the absorption rate does not
depend on the spatial and temporal variables \cite{DKH}. We note in this
context that the particle dynamics in an absorbing medium characterized by an
arbitrary dependence of the absorption rate on these variables can be studied
within the path-integral approach (see, e.g., Refs.~\cite{Path,MD}).

While the main purpose of Ref.~\cite{DKH} was to show both analytically and
numerically that the surviving and absorbing states exist, in this paper we
present a detailed analytical study of the distribution function of particles
subjected to superheavy-tailed noise. The paper is organized as follows. In
Section \ref{sec:Def} we describe the model and introduce the Langevin equation
driven by superheavy-tailed noise and the corresponding generalized
Fokker-Planck equation. The main results are obtained in Section \ref{sec:Gen}.
Here, using the theory of slowly varying functions, we confirm our previous
results, derive the probability distribution function of particles, and
establish a general connection between the noise characteristics and absorbtion
rate. In Section \ref{sec:Part}, for illustrative purposes, we consider a
special class of superheavy-tailed densities. Finally, our findings are
summarized in Section \ref{sec:Concl}.

\section{Definitions and basic equations}
\label{sec:Def}

It has been shown in Ref.~\cite{DHH1} that the probability density $P(x,t)$ of
the solution $x(t)$ of the first-order (overdamped) Langevin equation
\begin{equation}
    dx(t) = f(x(t),t)dt + d\eta(t)
\label{Langevin}
\end{equation}
satisfies the generalized Fokker-Planck equation
\begin{equation}
    \frac{\partial}{\partial t}P(x,t) = -\frac{\partial}{\partial x}
    f(x,t)P(x,t) + \mathcal{F}^{-1}\{P_{k}(t) \ln S_{k}\}.
    \label{FP1}
\end{equation}
Here, we interpret $x(t)$ [$x(0)=0$] as a particle position, $f(x,t)$ as a
deterministic force and $d\eta(t)$ as a noise term. The noise generating
process $\eta(t)$ [$\eta(0)=0$] is assumed to have independent and identically
distributed (for a given $dt$) increments $d\eta(t) = \eta(t+dt) - \eta(t)$. It
is also assumed that the solution of equation (\ref{FP1}) obeys the initial
condition $P(x,0) = \delta(x)$, where $\delta(x)$ is the Dirac $\delta$
function, and the normalization condition $\int_{-\infty}^{\infty} dxP(x,t)
=1$. The last term in (\ref{FP1}) contains the direct and inverse Fourier
transforms defined as $\mathcal{F} \{u_{k}\} \equiv u_{k} =
\int_{-\infty}^{\infty} dx\, e^{-ikx} u(x)$ and $\mathcal{F}^{-1} \{u_{k}\}
\equiv u(x) = (1/2\pi) \int_{-\infty}^ {\infty} dk\, e^{ikx} u_{k}$,
respectively. Specifically, $P_{k}(t) = \int_{-\infty}^{\infty}
dxe^{-ikx}P(x,t)$ is the characteristic function of $x(t)$, i.e., $P_{k}(t)=
\langle e^{-ikx(t)} \rangle$, where the angular brackets denote averaging over
the sample paths of the noise generating process $\eta(t)$, and $S_{k}= \langle
e^{-ik\eta(1)} \rangle$ is the characteristic function of the random variable
$\eta(1)$, i.e., noise generating process at $t=1$.

The main advantages of equation (\ref{FP1}) are that it (i) is valid for all
noise generating processes $\eta(t)$, (ii) accounts for the noise action in a
unified way, namely through the characteristic function $S_{k}$ of $\eta(1)$,
and (iii) reproduces all presently known Fokker-Planck equations associated
with the Langevin equation (\ref{Langevin}) \cite{DHH2}. The possible forms of
this equation, which are determined by the possible forms of the characteristic
function $S_{k}$, are restricted by the condition that the random variable
$\eta(1)$ is represented as an infinite sum of independent and identically
distributed increments $\Delta \eta(j\tau) = \eta(j\tau + \tau) - \eta(j\tau)$
when the split time $\tau$ tends to zero, i.e.,
\begin{equation}
    \eta(1) = \lim_{\tau \to 0} \sum_{j=0}^{ [1/\tau]-1}
    \Delta \eta(j\tau)
    \label{eta}
\end{equation}
($[1/\tau]$ is the integer part of $1/\tau$). For example, if the probability
density of the increments $\Delta \eta(j\tau)$ is given by
\begin{equation}
    p(\Delta \eta, \tau) = \frac{1}{a(\tau)}p\! \left(
    \frac{\Delta\eta}{a(\tau)}\right)\!,
    \label{tr}
\end{equation}
where the probability density $p(y)$ satisfies the condition $\lim_{\epsilon
\to 0} p(y/\epsilon) /\epsilon = \delta(y)$ and the scale function approaches
zero as $\tau \to 0$, then, depending on the asymptotic behavior of $p(y)$ at
$|y| \to \infty$, the ordinary or generalized central limit theorem can be
applied to determine $S_{k}$ \cite{DHH1,DHH2} (see also Sect.~\ref{sec:Con}).
It has been shown in particular that if $p(y)$ has heavy tails then $S_{k}$ is
the characteristic function of L\'{e}vy stable distributions and, as a
consequence, equation (\ref{FP1}) reduces to the fractional Fokker-Planck
equation. On the other hand, since \cite{DHH1}
\begin{equation}
    \ln S_{k} = \lim_{\tau \to 0} \frac{1}{\tau}[p_{ka(\tau)} - 1]
    \label{lnS}
\end{equation}
with $p_{ka(\tau)} = \int_{-\infty}^{\infty} dy\, e^{-ika(\tau) y} p(y)$, the
parameters of the last equation can be expressed through the asymptotic
behavior of the probability density $p(y)$ at $|y| \to \infty$ and the scale
function $a(\tau)$ at $\tau \to 0$ \cite{DHK}.

In this paper we focus on the study of the cumulative distribution function
$F(x,t) = \int_{-\infty}^{x} dxP(x,t)$ of particles subjected to
superheavy-tailed noises. These noises arise from the noise generating
processes characterized by the superheavy-tailed densities $p(y)$ whose
fractional moments $M_{\chi} = \int_{- \infty} ^{\infty}dy\, |y|^{\chi} p(y)$
are infinite for all $\chi>0$ (the normalization of $p(y)$ implies that
$M_{0}=1$). Next we restrict ourselves to the symmetric superheavy-tailed
densities with the following asymptotic behavior:
\begin{equation}
    p(y) \sim \frac{1}{y}h(y) \quad (y \to \infty),
    \label{as p}
\end{equation}
where a positive function $h(y)$ is assumed to be slowly varying at infinity,
i.e., $h(\mu y) \sim h(y)$ ($y \to \infty$) for all $\mu>0$. Since for all
slowly varying functions the condition $y^{\chi} h(y) \to \infty$ holds as
$y\to\infty$ \cite{BGT}, the probability densities characterized by the
asymptotic behavior (\ref{as p}) are indeed superheavy-tailed, i.e.,
$M_{\chi>0} = \infty$. It should be noted, however, that the slowly varying
functions $h(y)$ are not arbitrary because the condition $M_{0}=1$ implies that
$h(y) = o(1/ \ln y)$ as $y \to \infty$.

\section{General results}
\label{sec:Gen}

\subsection{Characteristic function \mbox{\boldmath $S_{k}$}}
\label{Char}

For finding the characteristic function $S_{k}$ of the random variable
$\eta(1)$ we use the basic relation (\ref{lnS}), which with the symmetry
property $p(-y)= p(y)$ and the normalization condition $2\int_{0}^{\infty}dy
p(y) =1$ can be rewritten as
\begin{equation}
    \ln S_{k} = -2 \lim_{\tau \to 0} \frac{1}{\tau}\, Y\! \left(
    \frac{1}{|k|a(\tau)} \right)\!,
    \label{lnS1}
\end{equation}
where
\begin{equation}
    Y(\lambda) = \int_{0}^{\infty} dy [1 - \cos(y/\lambda)]p(y)
    \label{Y}
\end{equation}
is a positive function of $\lambda$ that approaches zero as $\lambda \to
\infty$. In those cases when the indefinite integral $\int\! dyp(y)$ is known
(see, e.g., the example in Section \ref{sec:Part}) it is convenient to use an
alternative representation of $Y(\lambda)$:
\begin{equation}
    Y(\lambda) = \int_{0}^{\infty} dx\, \sin x \,
    \int_{\lambda x}^{\infty} dy\, p(y).
    \label{Y1}
\end{equation}

Since the function $h(y)$ in the asymptotic formula (\ref{as p}) is slowly
varying, $Y(\lambda)$ also varies slowly. Indeed, using (\ref{Y}), one can
easily verify that
\begin{eqnarray}
    Y(\mu\lambda) \!&=&\! \mu\lambda \int_{0}^{\infty} dx (1-\cos x)
    p(\mu\lambda x)
    \nonumber\\[6pt]
    &\sim& \! \int_{0}^{\infty} dx \frac{1-\cos x}{x} h(\mu\lambda x)
    \sim Y(\lambda) \quad
    \label{asrel1}
\end{eqnarray}
($\lambda \to \infty$). Therefore, if $k \neq 0$ then $Y(1/|k|a(\tau))$ in
(\ref{lnS1}) can be replaced by $Y(1/a(\tau))$ yielding $\ln S_{k} = -q$, where
\begin{equation}
    q = 2\lim_{\tau \to 0}\frac{1}{\tau}\,Y\bigg( \frac{1}
    {a(\tau)}\bigg)
    \label{q1}
\end{equation}
is a non-negative parameter that does not depend on $k$ (its physical meaning
will be discussed in Sect.~\ref{sec:SolF-P}). In contrast, if $k=0$ then
$p_{0}= \int_{-\infty}^{\infty} dyp(y) =1$ and so $\ln S_{0} =0$. Combining
these results and introducing the Kronecker delta $\delta_{k0}$, we obtain
\cite{DKH}
\begin{equation}
    \ln S_{k} = -q(1-\delta_{k0}).
    \label{lnS2}
\end{equation}
Thus, for all symmetric probability densities $p(y)$ with superheavy tails the
characteristic function of $\eta(1)$ has the form $S_{k} = e^{-q(1 -
\delta_{k0})}$. It should be noted that although $S_{k}$ differs from $e^{-q}$
only in one point $k=0$, the Kronecker delta cannot be neglected because it
provides a correct normalization of the distribution of $\eta(1)$ (see
Sect.~\ref{sec:Dis}).

\subsubsection{Representations of $q$ and $a(\tau)$}
\label{sec:q a}

As it follows from (\ref{q1}), the parameter $q$, which accounts for the
influence of superheavy-tailed noise on the system, depends on the asymptotic
behavior of the slowly varying function $Y(\lambda)$. In general, according to
(\ref{Y}), this behavior is controlled by a given probability density $p(y)$.
However, because $Y(\lambda)$ varies slowly, there is a possibility to write
$Y(\lambda)$ at $\lambda \to \infty$ in a quite general form. Such a
possibility is provided by the Karamata representation theorem \cite{BGT},
which states that for some $l>0$ and all $\lambda$ satisfying the condition
$\lambda \geq l$ every slowly varying function $L(\lambda)$ can be written in
the form
\begin{equation}
    L(\lambda) = g(\lambda) \exp\! \left( \int_{l}^{\lambda} du
    \frac{\epsilon(u)}{u} \right)\!,
    \label{L1}
\end{equation}
where $g(\lambda) \to g\in (0, \infty)$ and $\epsilon(\lambda) \to 0$ as
$\lambda \to \infty$. Since the defining property of $L(\lambda)$, i.e., $L(\mu
\lambda) \sim L(\lambda)$, involves only the asymptotic behavior of this
function, the representation (\ref{L1}) is essentially non-unique. Using this
fact, it can be shown \cite{BGT} that $L(\lambda)$ at $\lambda \to \infty$ can
always be represented in a simpler form
\begin{equation}
    L(\lambda) \sim g \exp\! \left( \int_{\ln l}^{\ln \lambda} dv
    \epsilon(e^{v}) \right)\!.
    \label{L2}
\end{equation}

Thus, assuming that $\int\! dv \epsilon(e^{v}) = -\Phi(v)$, for the slowly
varying function $Y(\lambda)$ we obtain
\begin{equation}
    Y(\lambda) \sim c \exp[-\Phi(\ln \lambda)]
    \label{Y2}
\end{equation}
($\lambda \to \infty$) with $c=g\exp[\Phi(\ln l)]$. An explicit form of the
function $\Phi(v)$ depends on $p(y)$, but since $Y(\lambda)$ tends to zero when
$\lambda$ increases, the condition $\Phi(v) \to \infty$ as $v \to \infty$ must
hold in all cases. Using (\ref{q1}) and (\ref{Y2}), for the parameter $q$ we
find the desired general representation
\begin{equation}
    q = \frac{2c}{r},
    \label{q1b}
\end{equation}
where $r$ is a time scale parameter defined as
\begin{equation}
    r = \lim_{\tau \to 0} \tau \exp\!\left[\Phi\!\left(\ln \frac{1}
    {a(\tau)}\right)\right]\!.
    \label{r}
\end{equation}
Depending on how fast the function $a(\tau)$ approaches zero as $\tau \to 0$,
the parameter $r$ can take the values from the whole interval $[0, \infty]$. If
$a(\tau)$ tends to zero so rapidly that $r=\infty$ then such noise does not
affect the system at all ($q=0$). In the opposite case, when $r=0$, the noise
action is so strong that the system reaches the final state at $t=0^{+}$, i.e.,
immediately ($q=\infty$).

In what follows we are interested in a non-trivial action of superheavy-tailed
noise. This case is specified by the condition $0<r<\infty$ and occurs only if
the scale function $a(\tau)$ has a proper asymptotic behavior at $\tau \to 0$.
In order to find this behavior, we assume that for large enough $\varphi$ there
exists the inverse function $v=\Phi^{-1}(\varphi)$ of the function $\varphi =
\Phi(v)$ (since $\Phi(v) \to \infty$ as $v \to \infty$, in this case $\Phi^{-1}
(\varphi) \to \infty$ as $\varphi \to \infty$). Then the limit in (\ref{r}) is
equal to a given value $r$ if the asymptotic behavior of $a(\tau)$ at $\tau \to
0$ is given by
\begin{equation}
    a(\tau) \sim \exp\!\left[-\Phi^{-1}\!\left(\ln \frac{r}
    {\tau}\right)\right]\!.
    \label{a}
\end{equation}
We note that the asymptotic formula (\ref{a}) is not unique because, due to the
presence of the logarithmic function in (\ref{r}), any scale function of the
form $a(\tau)b(\tau)$ with $b(\tau)$ satisfying the condition
\begin{equation}
    \tau \exp \left\{ \Phi\left[\Phi^{-1}\! \left(\ln \frac{r}
    {\tau}\right) - \ln b(\tau) \right]\right\} \sim r
    \label{b}
\end{equation}
($\tau \to 0$) also leads to the same parameter $r$. It is important to
emphasize, however, that this lack of uniqueness is of no importance because
the parameters $c$ and $r$ completely characterize the noise action. Therefore,
at fixed $c$ and $r$ the Langevin equation (\ref{Langevin}) driven by
superheavy-tailed noise and the corresponding Fokker-Planck equation
(\ref{FP1}) are well defined and the function $b(\tau)$ can always be chosen as
$b(\tau)=1$ without loss of generality.

\subsection{Distribution function of \mbox{\boldmath $\eta(1)$}}
\label{sec:Dis}

In general, using (\ref{FP1}) and (\ref{lnS2}), it is possible to establish the
main features of the probability density $P(x,t)$ arising from the action of
superheavy-tailed noise \cite{DKH}. However, because the random variable
$\eta(1)$ is responsible for all these features, it is reasonable to study the
statistical properties of $\eta(1)$ in more detail. This will also permit us to
obtain the probability distribution function of the particle position,
$F(x,t)$, in a consistent way.

In order to gain more insight into the properties of the random variable
$\eta(1)$, we calculate its distribution function $F(\eta)$. To this end it is
convenient to temporarily replace the Kronecker delta $\delta_{k0}$ in the
characteristic function $S_{k} = e^{-q(1-\delta_ {k0})}$ of $\eta(1)$ by a
symmetric and smooth function of $k$, $\Delta_{k}(\kappa)$, which depends on a
positive parameter $\kappa$ in such a way that $\lim_{\kappa \to 0} \Delta_{k}
(\kappa) = \delta_{k0}$. We assume that $S_{k}(\kappa) = e^{-q[1-
\Delta_{k}(\kappa)]}$ is the characteristic function of the random variable
$\eta(1; \kappa)$ whose probability density $S(\eta; \kappa)$ and the
corresponding distribution function $F(\eta; \kappa)$ are given by $S(\eta;
\kappa) =(1/2\pi) \int_{-\infty} ^{\infty} dk\, e^{ik\eta} S_{k}(\kappa)$ and
$F(\eta; \kappa) = \int_{-\infty}^{\eta} d\eta S(\eta; \kappa)$, respectively.
If the distribution function $F(\eta; \kappa)$ is known then the desired
distribution function can be determined as $F(\eta) = \lim_{\kappa \to 0}
F(\eta; \kappa)$.

The choice of the function $\Delta_{k}(\kappa)$ is, of course, not unique. For
example, it can be chosen as
\begin{equation}
    \Delta_{k}(\kappa) = \frac{1}{q} \ln\! \left[1 +
    (e^{q}-1)\, \mathrm{sech}\! \left( \frac{\pi k}
    {2\kappa}\right)\right]
    \label{Delta}
\end{equation}
($\mathrm{sech}\,x = 1/\cosh x$) or
\begin{equation}
    \Delta_{k}(\kappa) = \frac{1}{q} \ln\! \left[1 +
    (e^{q}-1)\,e^{-|k|/\kappa} \right]\!.
    \label{Delta2}
\end{equation}
However, in the limit $\kappa \to 0$ the explicit form of $\Delta_{k}(\kappa)$
is of no importance. Therefore, without loss of generality, we consider
$\Delta_{k}(\kappa)$ from (\ref{Delta}) for which the characteristic function
of $\eta(1; \kappa)$ has the form
\begin{equation}
    S_{k}(\kappa) = e^{-q} + (1 - e^{-q})\, \mathrm{sech}\!
    \left( \frac{\pi k} {2\kappa}\right)\!.
    \label{Sk}
\end{equation}
The probability density and the distribution function that correspond to this
characteristic function are given by
\begin{equation}
    S(\eta; \kappa) = e^{-q}\delta(\eta) + (1 - e^{-q})\,
    \frac{\kappa}{\pi}\, \mathrm{sech}\,\kappa\eta
    \label{S}
\end{equation}
and
\begin{equation}
    F(\eta; \kappa) = e^{-q}F_{\rm{d}}(\eta) + (1 - e^{-q})
    \widetilde{F}(\kappa\eta),
    \label{F1}
\end{equation}
respectively. Here,
\begin{equation}
    F_{\rm{d}}(\eta) = \left\{ \begin{array}{ll}
    0, & \eta < 0
    \\[3pt]
    1, & \eta \geq 0
    \end{array}
    \right.
    \label{F}
\end{equation}
is the step function describing the degenerate distribution localized in the
point $\eta =0$, and
\begin{equation}
    \widetilde{F}(\kappa\eta) = \frac{2}{\pi}
    \arctan (e^{\kappa\eta})
    \label{F2}
\end{equation}
is the distribution function which describes the hyperbolic secant
distribution.

According to (\ref{F1}), the distribution of the random variable $\eta(1;
\kappa)$ is the mixing of the degenerate and hyperbolic secant distributions.
In contrast to the former, which  does not depend on the parameter $\kappa$,
the latter strongly depends on this parameter. In particular, the maximum
height of the distribution, $d \widetilde{F}(\kappa\eta)/d\eta|_{\eta =0}$, and
the root square of the variance, which characterizes the width of this
distribution, are equal to $\kappa/\pi$ and $\pi/(2\kappa)$, respectively. It
is clear, therefore, that the probability density $d \widetilde{F} (\kappa\eta)
/d\eta$ of the hyperbolic secant distribution tends to zero as $\kappa \to 0$.
On the other hand, as it follows from (\ref{F2}), this occurs in such a way
that $\widetilde{F} (-\infty) =0$ and $\widetilde{F} (\infty) =1$ for all
$\kappa>0$. Assuming that at $\kappa \to 0$ these conditions hold as well, for
the limiting distribution function $F_{\rm{l}}(\eta) = \lim_{\kappa \to 0}
\widetilde{F} (\kappa \eta)$ we obtain
\begin{equation}
    F_{\rm{l}}(\eta) =  \left\{ \begin{array}{ll}
    0, & \eta = -\infty
    \\[3pt]
    1/2, & |\eta| < \infty
    \\[3pt]
    1, & \eta = \infty.
    \end{array}
    \right.
    \label{Fs}
\end{equation}
This particular distribution function describes a normalized random variable
whose probability density equals zero for all $|\eta|<\infty$. From a formal
point of view, it can also be considered as a discrete variable that takes only
two values, $\eta = -\infty$ and $\eta = \infty$, with probability 1/2 each.
Thus, the probability distribution of $\eta(1)$ characterized by the
distribution function $F(\eta) = \lim_{\kappa \to 0} F(\eta; \kappa)$ is the
mixing of the degenerate distribution (\ref{F}) taken with the weight $e^{-q}$
and the limiting distribution (\ref{Fs}) taken with the weight $1-e^{-q}$,
i.e.,
\begin{equation}
    F(\eta) = e^{-q}F_{\rm{d}}(\eta) + (1 - e^{-q})
    F_{\rm{l}}(\eta).
    \label{F3}
\end{equation}

\subsubsection{Connection with limit theorems}
\label{sec:Con}

Representing the increments $\Delta \eta (j\tau)$ of the noise generating
process $\eta(t)$ as $\Delta \eta (j\tau) = y_{j}/c_{n}$ with $c_{n} =
1/a(\tau)$ and $n = [1/\tau] +1$, from (\ref{eta}) one obtains
\begin{equation}
    \eta(1) = \lim_{n \to \infty} \sum_{j=1}^{ n} \frac{y_{j}}{c_{n}},
    \label{eta2}
\end{equation}
where $y_{j}$ are the independent random variables distributed with the same
probability density $p(y)$. We note that in probability theory the partial sums
of a more general form, i.e., $\sum_{j=1}^{ n} y_{j}/c_{n} - d_{n}$, is usually
considered. But in our situation $d_{n}=0$ because a class of probability
densities $p(y)$ is restricted by the condition $\lim_{\epsilon \to 0}
p(y/\epsilon)/ \epsilon = \delta(y)$. If the variance of $p(y)$ is finite then,
according to the central limit theorem \cite{Fel}, $c_{n} \propto n^{1/2}$ and
the distribution of $\eta(1)$ is Gaussian. In this case $a(\tau) \propto
\tau^{1/2}$ and for all such $p(y)$ the generalized Fokker-Planck equation
(\ref{FP1}) reduces to the ordinary one \cite{DHK}. In contrast, if $p(y)$ with
heavy tails belongs to the domain of normal attraction of a given stable
distribution characterized by an index of stability $\alpha \in (0,2)$ then, as
the generalized central limit theorem suggests \cite{Fel}, $n \propto n^{1/
\alpha}$ and $\eta(1)$ is described by this stable distribution. It has been
shown \cite{DHK} that in this case $a(\tau) \propto \tau^{1/ \alpha}$ [in
special situations, there also exist two different forms of the scale function
$a(\tau)$] and equation (\ref{FP1}) reduces to the fractional Fokker-Planck
equation.

Finally, the above derived results show that for all symmetric
superheavy-tailed densities $p(y)$ characterized by the asymptotic behavior
(\ref{as p}) the distribution function of the random variable $\eta(1)$ has the
form (\ref{F3}). It is important to emphasize, however, that in contrast to
densities with finite variance and densities with heavy tails there is no
universal scale function $a(\tau)$ for superheavy-tailed $p(y)$. On the
contrary, as is clear from (\ref{Y1}), (\ref{Y2}) and (\ref{a}), in this case
different $p(y)$ lead in general to different $a(\tau)$. This is in accordance
with the well-known fact \cite{Fel} that only stable distributions have domains
of attraction. Thus, the result (\ref{F3}) can be interpreted as follows: For
every symmetric superheavy-tailed $p(y)$ the scale function $a(\tau)$ can
always be chosen so that the distribution function of $\eta(1)$ is given by
(\ref{F3}) with $q \in (0,\infty)$.

\subsection{Solution of the generalized Fokker-Planck equation}
\label{sec:SolF-P}

Our next step is to solve the generalized Fokker-Planck equation (\ref{FP1})
for the characteristic function $S_{k} = e^{-q(1-\delta_ {k0})}$ that
corresponds to superheavy-tailed noise. To this end we first solve this
equation for the characteristic function (\ref{Sk}) and then take the limit
$\kappa \to 0$. In order to clarify our approach, let us designate the solution
of equation (\ref{FP1}) in which $S_{k}$ is replaced by $S_{k}(\kappa)$ as
$P(x,t;\kappa)$. Then we decompose it as $P(x,t;\kappa) = \mathcal{P} (x,t) +
\mathcal{A}(x,t;\kappa)$, where the terms $\mathcal{P} (x,t)$ and
$\mathcal{A}(x,t;\kappa)$ are governed by the equations
\begin{eqnarray}
    \frac{\partial}{\partial t}\mathcal{P}(x,t) \!&=&\!
    -\frac{\partial}{\partial x}f(x,t)\mathcal{P}(x,t) -
    q\mathcal{P}(x,t),
    \label{FP2}
    \\[6pt]
    \frac{\partial}{\partial t}\mathcal{A}(x,t;\kappa) \!&=&\!
    -\frac{\partial}{\partial x}f(x,t)\mathcal{A}(x,t;\kappa) -
    q\mathcal{A}(x,t;\kappa)
    \nonumber\\[3pt]
    &&\! +\, q\mathcal{F}^{-1} \{[\mathcal{P}_{k}(t)
    + \mathcal{A}_{k}(t;\kappa)] \Delta_{k}(\kappa)\} \qquad\quad
    \label{A}
\end{eqnarray}
and satisfy the initial conditions $\mathcal{P}(x,0) = \delta(x)$ and
\linebreak $\mathcal{A} (x,0;\kappa)=0$. We also assume that the normalization
condition for $P(x,t;\kappa)$, i.e., $P_{k}(t;\kappa)|_{k=0} =
\mathcal{P}_{0}(t) + \mathcal{A}_{0}(t;\kappa)= 1$, holds for all $t$ and
$\kappa$. Then, with these assumptions, the solution of equation (\ref{FP1})
can be represented as follows:
\begin{equation}
    P(x,t)= \mathcal{P}(x,t) + \lim_{\kappa \to 0}
    \mathcal{A}(x,t;\kappa).
    \label{repr}
\end{equation}

An explicit form of the first term in the right-hand side of (\ref{repr}) can
easily be determined from equation (\ref{FP2}). Indeed, assuming that
$\mathcal{P}(x,t) = e^{-qt} \mathcal{R}(x,t)$, this equation reduces to
\begin{equation}
    \frac{\partial}{\partial t}\mathcal{R}(x,t) =
    -\frac{\partial}{\partial x}f(x,t)\mathcal{R}(x,t).
    \label{R}
\end{equation}
Its general solution is written as $\mathcal{R}(x,t) = \Phi[x - z(t)]$, where
$\Phi(x)$ is an arbitrary function and $z(t)$ is the solution of the equation
$\dot{z}(t) = f(z(t),t)$ (we assume that $z(0)=0$ and $|z(t)|<\infty$ for all
finite $t$). Using the initial condition $\mathcal {R}(x,0) = \delta(x)$, one
finds $\Phi(x) = \delta(x)$ and so
\begin{equation}
    \mathcal{P}(x,t) = e^{-qt}\delta[x-z(t)].
    \label{sol1}
\end{equation}
As it follows from this result, the normalization condition for
$\mathcal{P}(x,t)$, $\mathcal{P}_{0}(t) = \int_{-\infty}^{\infty}
dx\mathcal{P}(x,t)$, reads $\mathcal{P}_{0}(t) = e^{-qt}$.

According to (\ref{A}) and (\ref{sol1}), the term $\mathcal{A} (x,t;\kappa)$
obeys a closed integro-differential equation
\begin{eqnarray}
    \frac{\partial}{\partial t}\mathcal{A}(x,t;\kappa) \!&=&\!
    -\frac{\partial}{\partial x}f(x,t)\mathcal{A}(x,t;\kappa) -
    q\mathcal{A}(x,t;\kappa)
    \nonumber\\[3pt]
    &&\! +\, q \int_{-\infty}^{\infty} dy \Delta(x-y;\kappa)
    \mathcal{A}(y,t;\kappa)
    \nonumber\\[3pt]
    &&\! +\, q e^{-qt} \Delta(x-z(t);\kappa),
    \label{A1}
\end{eqnarray}
where $\Delta(x;\kappa) = (1/2\pi) \int_{-\infty}^{\infty} dk\, e^{ikx}
\Delta_{k} (\kappa)$ is the kernel function. Using the fact that
$\int_{-\infty}^ {\infty}dx \Delta (x,\kappa) =1$, which follows from the
condition $\Delta_{0} (\kappa) = 1$ [see (\ref{Delta})], and the integral
representation of the $\delta$ function $\delta(x) = (1/2\pi)\times \int_{
-\infty} ^{\infty}dke^{ikx}$, from equation (\ref{A1}) we obtain a simple
equation $\partial \mathcal{A}_{0}(t;\kappa)/\partial t  = qe^{-qt}$ for
$\mathcal{A}_{0} (t;\kappa) = \int_{-\infty}^{\infty} dx \mathcal{A}
(x,t;\kappa)$. The solution of this equation satisfying the initial condition
$\mathcal{A}_{0}(0;\kappa) =0$ yields $\mathcal{A}_{0}(t;\kappa) = 1-e^{-qt}$.
Therefore, like $\mathcal{P} (x,t)$, the normalization condi\-tion of
$\mathcal{A} (x,t;\kappa)$, i.e., $\mathcal{A}_{0}(t;\kappa) = 1-e^{-qt}$, does
not depend on the parameter $\kappa$.

Because of its complexity, we are not able to solve equation (\ref{A1}) for an
arbitrary $\kappa$. However, in the case of our interest, when $\kappa$
approaches zero, the dependence of the solution of this equation on $\kappa$
can easily be found. The key point is that the kernel function $\Delta
(x,\kappa)$ at $|x|\kappa \ll 1$ can be approximated as $\Delta(x;\kappa) = \xi
\kappa$ with
\begin{equation}
    \xi = \frac{2}{\pi^{2}q} \int_{0}^{\infty} dy \ln [1 +
    (e^{q}-1)\, \mathrm{sech}\, y].
    \label{rel1}
\end{equation}
In this approximation, the integral term in (\ref{A1}) equals $q\xi (1 -
e^{-qt})\kappa$ and $\mathcal{A}(x,t;\kappa)$ at $|x - z(t)|\kappa\ll 1$ is
governed by a linear differential equation
\begin{equation}
    \frac{\partial}{\partial t}\mathcal{A}(x,t;\kappa) =
    -\frac{\partial}{\partial x}f(x,t)\mathcal{A}(x,t;\kappa) -
    q\mathcal{A}(x,t;\kappa) + q\xi\kappa.
    \label{A2}
\end{equation}
Its general solution is written as $\mathcal{A} (x,t;\kappa) = \mathcal{A}_
{\rm{p}} (x,t;\kappa) + \mathcal{A}_{\rm{g}} (x,t;\kappa)$, where $\mathcal{A}_
{\rm{p}} (x,t;\kappa) \propto \kappa$ is a particular solution of this equation
and $\mathcal{A}_{\rm{g}} (x,t;\kappa)$ is the general solution of the
corresponding homogeneous equation (when $\kappa =0$). Since $\mathcal{A}
(x,0;\kappa) =0$, the latter is also proportional to $\kappa$ and so
$\mathcal{A} (x,t;\kappa) \propto \kappa$. Put differently, the term
$\mathcal{A} (x,t;\kappa)$ tends to zero linearly as $\kappa \to 0$. In
particular, if the external force is a constant, i.e., $f(x,t) =f$, then
$\mathcal{A} (x,t;\kappa) = \xi(1-e^{-qt}) \kappa$ at $|x - ft| \kappa \ll 1$.
We note also that while $\mathcal{A} (x,t;\kappa)$ approaches zero in the limit
$\kappa \to 0$, the spatial region of localization of $\mathcal{A} (x,t;
\kappa)$, which is determined by the condition $|x - z(t)| \sim 1/\kappa$,
tends to infinity keeping the normalization condition of $\mathcal{A}
(x,t;\kappa)$ fixed.

Thus, according to (\ref{repr}), (\ref{sol1}) and the above properties of
$\mathcal{A} (x,t;\kappa)$ at $\kappa \to 0$, the distribution function $F(x,t)
= \int_{-\infty}^{x}dx P(x,t)$ ($t<\infty$) of particles driven by
superheavy-tailed noise has the form
\begin{equation}
    F(x,t) = e^{-qt}F_{\rm{d}}[x-z(t)] + (1 - e^{-qt})
    F_{\rm{l}}(x).
    \label{F(x,t)}
\end{equation}
It shows that this noise generates two time-dependent probabilistic states of
each particle. The first one is described by the degenerate distribution
function $F_{\rm{d}} [x-z(t)]$ and is realized with the probability
$\mathcal{P}_{ 0} (t) = e^{-qt}$. Since the particle trajectory $z(t)$ is not
influenced by the noise [see (\ref{sol1})], following the terminology of
\cite{DKH} we call this state the surviving state. The second state is
associated with the limiting distribution function $F_{\rm{l}}(x)$ and is
realized with the probability $\mathcal{A}_{0}(t) = 1- e^{-qt}$. In this state
the probability to find the particle in any finite interval equals zero. The
transition to this state implies that the particle jumps to plus or minus
infinity and, in fact, it is excluded from future consideration. Therefore,
like in \cite{DKH}, we refer to this state as the absorbing state. Because of
these features, superheavy-tailed noise acts on particles as an absorbing
medium characterized by the rate of absorption $q$. Hence, the Langevin
equation (\ref{Langevin}) driven by this noise and the corresponding
generalized Fokker-Planck equation (\ref{FP1}) describe the overdamped motion
of a particle in an absorbing medium. Moreover, one expects that these
equations can also be used to describe some other processes whose duration is
random.

\subsection{Compound noise}
\label{sec:Comp}

According to the above results, superheavy-tailed noise is characterized by two
main features which at first glance seem to be contradictory. On the one hand,
this noise is so strong that the transition into the absorbing state occurs so
that a particle is immediately transferred to infinity. But, on the other hand,
before this transition the noise does not affect the particle position. In
other words, before absorption which is random in time the particle motion is
deterministic and is governed by the equation of motion $\dot{z}(t) = f(z(t),
t)$ [see (\ref{sol1})]. From a mathematical point of view, these features
result from both superheavy tails of the probability density $p(y)$ and
appropriate choice of the scale function $a(\tau)$.

The case when before absorption the particle motion is random can also be
incorporated into the proposed approach. For this purpose we consider the
compound noise generating process $\eta(t) = \eta_{1}(t) + \eta_{2}(t)$, where
$\eta_{1}(t)$ generates superheavy-tailed noise, which models an absorbing
medium, and $\eta_{2}(t)$ generates any other noise (e.g., L\'{e}vy stable
noise), which causes the random motion of a particle. Assuming that these
random processes are independent, we obtain $S_{k} = S_{1k}S_{2k}$, where
$S_{1k}$ and $S_{2k}$ are the characteristic functions of $\eta_{1}(1)$ and
$\eta_{2}(1)$, respectively. Then, proceeding as before, we replace $S_{1k}$ by
the characteristic function (\ref{Sk}) and write the solution of equation
(\ref{FP1}) at $\kappa \ne 0$ as $P(x,t;\kappa) = \mathcal{P} (x,t) +
\mathcal{A}(x,t; \kappa)$, where $\mathcal{P} (x,t)$ is the solution of the
equation
\begin{eqnarray}
    \frac{\partial}{\partial t}\mathcal{P}(x,t) \!&=&\!
    -\frac{\partial}{\partial x}f(x,t)\mathcal{P}(x,t) -
    q\mathcal{P}(x,t)
    \nonumber\\[3pt]
    &&\! +\, \mathcal{F}^{-1} \{\mathcal{P}_{k}(t)
    \ln S_{2k}\}
    \label{FP3}
\end{eqnarray}
satisfying the initial condition $\mathcal{P}(x,0) = \delta(x)$. This solution
can be written in the form $\mathcal{P} (x,t) = e^{-qt}\mathcal{W}(x,t)$, where
the probability density $\mathcal{W}(x,t)$ is governed by the generalized
Fokker-Planck equation
\begin{equation}
    \frac{\partial}{\partial t}\mathcal{W}(x,t) =
    -\frac{\partial}{\partial x}f(x,t)\mathcal{W}(x,t)
    + \mathcal{F}^{-1} \{\mathcal{W}_{k}(t)\ln S_{2k}\}
    \label{FP4}
\end{equation}
and obeys the initial condition $\mathcal{W}(x,0) = \delta(x)$ and the
normalization condition $\mathcal{W}_{0}(t) =1$. It is also not difficult to
verify that, like in the previous case, $\mathcal{A}_{0}(t;\kappa) =1 -
e^{-qt}$ and $\mathcal{A}(x,t;\kappa) \to 0$ as $\kappa \to 0$. Therefore, for
all finite times the distribution function of particles subjected to this
compound noise takes the form
\begin{equation}
    F(x,t) = e^{-qt}\int_{-\infty}^{x}dx \mathcal{W}(x,t) +
    (1 - e^{-qt}) F_{\rm{l}}(x).
    \label{F(x,t)2}
\end{equation}

Using the fact that superheavy-tailed noise plays the role of an absorbing
medium, the above result can be interpreted as the distribution function of
particles whose random motion before absorption is described by the following
Langevin equation: $dx(t) = f(x(t),t)dt + d\eta_{2}(t)$. The second term in the
right-hand side of (\ref{F(x,t)2}) relates to the absorbing state and does not
depend on the character of the particle motion. In contrast, the particle
dynamics strongly influences the first term which describes the surviving
state. As is clear from (\ref{FP4}) and (\ref{F(x,t)2}), this influence is
determined by the deterministic force $f(x,t)$ and the characteristic function
$S_{2k}$ of $\eta_{2}(1)$. If $\eta_{2}(t)=0$, i.e., the particle dynamics
before absorption is deterministic, then $S_{2k}=1$, $\mathcal{W}(x,t) =
\delta[x - z(t)]$ and so the distribution function (\ref{F(x,t)2}) reduces to
(\ref{F(x,t)}). The random motion of particles occurs if the noise generating
process $\eta_{2}(t)$ depends on time. As it was discussed in
Sect.~\ref{sec:Con}, in this case the possible distributions of $\eta_{2}(1)$
follow from the fact that $\eta_{2}(1)$ is represented as an infinite sum of
independent and identically distributed increments. For example, if the
distribution of the increments $\Delta \eta_{2}(j\tau)$ is symmetric and has
heavy tails then the random variable $\eta_{2}(1)$ is distributed with the
symmetric L\'{e}vy stable distribution whose characteristic function is given
by $S_{2k} = \exp(-\gamma |k|^{\alpha})$, where $\alpha$ and $\gamma$ are the
index of stability and scale parameter, respectively. Defining the Riesz
derivative as $\partial^{\alpha}h(x)/\partial |x|^{\alpha} = -\mathcal{F}^{-1}
\{ |k|^{\alpha} h_{k} \}$ \cite{SKM}, in this particular case equation
(\ref{FP4}) reduces to the fractional Fokker-Planck equation
\cite{JMF,Dit,SLDYL,BS,DS,DHH1}
\begin{equation}
    \frac{\partial}{\partial t}\mathcal{W}(x,t) =
    -\frac{\partial} {\partial x}f(x,t)\mathcal{W}(x,t) +
    \gamma \frac{\partial^{\alpha}}{\partial |x|^{\alpha}}
    \mathcal{W}(x,t),
    \label{FFP}
\end{equation}
which together with the distribution function (\ref{F(x,t)2}) describes
L\'{e}vy flights (if $0<\alpha<2$) and Brownian motion of particles (if
$\alpha=2$) in an absorbing medium.

Equation (\ref{FFP}) can be solved analytically in some simple cases. In
particular, if $f(x,t)$ is a linear restoring force, i.e., $f(x,t) = -bx$
($b>0$), then the solution of this equation is the L\'{e}vy stable density
\cite{JMF}
\begin{equation}
    \mathcal{W}(x,t) = \mathcal{F}^{-1} \{\exp (-c^{2}(t)
    |k|^{\alpha}) \}
    \label{W1}
\end{equation}
characterized by the same index of stability and the time-dependent scale
parameter $c^{2}(t) = \gamma (1- e^{-\alpha bt})/\alpha b$. At $\alpha = 2$ it
reduces to the Gaussian density
\begin{equation}
    \mathcal{W}(x,t) = \frac{1}{2\sqrt{\pi}c(t)}\exp\!
    \left(- \frac{x^{2}}{4c^{2}(t)}\right)\!,
    \label{W2}
\end{equation}
and the distribution function (\ref{F(x,t)2}) for Brownian particles in an
absorbing medium takes the form
\begin{equation}
    F(x,t) = \frac{e^{-qt}}{2}\, {\rm{erfc}}\!
    \left(- \frac{x}{2c(t)} \right) +
    (1 - e^{-qt}) F_{\rm{l}}(x),
    \label{F(x,t)3}
\end{equation}
where ${\rm{erfc}}(x) = (2/\sqrt{\pi}) \int_{x}^{\infty}dye^{-y^{2}}$ is the
complementary error function.

\section{Particular class of superheavy-tailed distributions}
\label{sec:Part}

As is shown in the previous section, the distribution of particles subjected to
superheavy-tailed noise is the same as their distribution in an absorbing
medium characterized by the rate of absorption $q$. According to (\ref{q1}),
this rate is determined by both the asymptotic behavior of the probability
density $p(y)$ at $y \to \infty$ and the asymptotic behavior of the scale
function $a(\tau)$ at $\tau \to 0$. Since $p(y)$ is assumed to be given, the
noise action is non-trivial, i.e., $0<q<\infty$, if the asymptotic behavior of
$a(\tau)$ is given by (\ref{a}). It should be noted that finding the scale
function $a(\tau)$ at small $\tau$ is important not only for determining the
absorption rate $q$, but also for the numerical simulations of the Langevin
equation \cite{DKH}.

In order to illustrate how our general results can be used for the
determination of $q$ and $a(\tau)$ at $\tau \to 0$, we consider a
two-parametric class of superheavy-tailed distributions that is described by
the probability density
\begin{equation}
    p(y) = \frac{(\nu - 1)(\ln s)^{\nu -1}}{2(s+|y|)
    [\ln (s+|y|)]^{\nu}}.
    \label{p2}
\end{equation}
Since $p(y)$ is assumed to be normalized and non-negative, the parameters $\nu$
and $s$ must satisfy the conditions $\nu>1$ and $s>1$. According to (\ref{p2}),
the probability density $p(y)$ is symmetric, has a single maximum located at
$y=0$, and its asymptotic behavior at $y \to \infty$ is given by (\ref{as p})
with
\begin{equation}
    h(y) \sim \frac{(\nu - 1)(\ln s)^{\nu -1}}
    {2(\ln y)^{\nu}}.
    \label{as h}
\end{equation}

For the probability density (\ref{p2}) the inner integral in (\ref{Y1}) can
easily be calculated yielding
\begin{equation}
    Y(\lambda) = \frac{(\ln s)^{\nu -1}}{2}\int_{0}^{\infty} dx\,
    \frac{\sin x}{[\ln (s+\lambda x)]^{\nu -1}}.
    \label{Y3}
\end{equation}
To find the leading term of the asymptotic expansion of $Y(\lambda)$ as
$\lambda \to \infty$, we use the identity
\begin{equation}
    \frac{1-e^{- \sigma\rho}}{ \sigma} = \int_{0}^{\rho} du e^{-\sigma u}
    \label{rel a}
\end{equation}
with $\sigma= [\ln (s+\lambda x)]^{\nu -1}$ and $\rho = (\ln \lambda)^{2-\nu}$.
Introducing the variable of integration $z=u/\rho$ and taking into account that
$\sigma \rho \sim \ln(x\lambda) \sim \ln \lambda$ if $\lambda \to \infty$ and
$0<x<\infty$, from (\ref{rel a}) we obtain
\begin{equation}
    \frac{1}{ [\ln (s+\lambda x)]^{\nu -1}} \sim
    \frac{1}{ (\ln \lambda)^{\nu -2}} \int_{0}^{1} dz
    \frac{1}{(\lambda x)^{z}}.
    \label{rel b}
\end{equation}
Then, substituting (\ref{rel b}) into (\ref{Y3}) and using the integral formula
$\int_{0}^{\infty}dx x^{-z}\sin x = \Gamma(1-z) \cos(\pi z/2)$, where
$\Gamma(z)$ is the gamma function, we arrive to
\begin{equation}
    Y(\lambda) \sim \frac{(\ln s)^{\nu -1}}{2(\ln \lambda)^{\nu -2}}
    \int_{0}^{1}dz e^{-z\ln\lambda} \Gamma(1-z) \cos \frac{\pi z}{2}.
    \label{Y4}
\end{equation}
Since $\lambda \to \infty$, the main contribution to the integral in (\ref{Y4})
comes from a small vicinity of the lower limit of integration. Making use of
this fact, one can easily verify that this integral is asymptotically equal to
$1/\ln \lambda$, and so the asymptotic formula (\ref{Y4}) takes the form
\begin{equation}
    Y(\lambda) \sim \frac{1}{2}\! \left( \frac{\ln s}{\ln \lambda}
    \right)^{\! \nu -1}.
    \label{Y5}
\end{equation}

We are now in a position to find the asymptotic representation of the functions
$\Phi(v)$ and $\Phi^{-1} (\varphi)$ in the reference case [we remind that
according to (\ref{q1b})--(\ref{a}) these functions determine the absorption
rate $q$ and the scale function $a(\tau)$]. By comparing (\ref{Y5}) with
(\ref{Y2}), one obtains $c = (\ln s)^{\nu-1}/2$ and $\Phi(v) \sim (\nu-1) \ln
v$ ($v \to \infty$). The first result leads to the absorption rate
\begin{equation}
    q = \frac{(\ln s)^{\nu -1}}{r},
    \label{q2}
\end{equation}
and the second permits us to find an explicit asymptotic formula for the
inverse function: $\Phi^{-1}(\varphi) \sim \exp[\varphi /(\nu-1)]$ ($\varphi
\to \infty$). Using this formula, from (\ref{a}) we get
\begin{equation}
    a(\tau) \sim \exp\! \left[ -\exp\! \left( \frac{1}{\nu -1}
    \ln \frac{r}{\tau}\right)\right]
    \label{a2}
\end{equation}
as $\tau \to 0$. Thus, the noise action is specified not only by the parameters
$\nu$ and $s$, which characterize the superheavy-tailed probability density
(\ref{p2}), but also by the parameter $r$, which characterizes the asymptotic
behavior (\ref{a2}) of the scale function $a(\tau)$. We note also that at
$\nu=2$ the formulas (\ref{q2}) and (\ref{a2}) are reduced to those derived in
Ref.~\cite{DKH}, i.e., $q=(\ln s)/r$ and $a(\tau) \sim \exp(-r/\tau)$.

We complete our analysis by comparing the sample paths of the solutions $x(t)$
of the Langevin equation (\ref{Langevin}) driven by different noises. If the
transition probability density of the noise generating process is given by
(\ref{tr}) then, depending on the probability density $p(y)$, the sample paths
can show three qualitatively different behaviors. Specifically, if the variance
of $p(y)$ is finite then the sample paths of $x(t)$ are random and continuous
(e.g., as in the case of the Wiener process). In contrast, the sample paths
which correspond to the probability density $p(y)$ with heavy tails are random
and discontinuous (e.g., as in the case of the L\'{e}vy process). Finally, if
$p(y)$ has superheavy tails then the corresponding noise does not influence
$x(t)$ up to a random time $t_{\rm{tr}}$ whose average value, $\overline{
t_{\rm{tr}}} = \int_{0} ^{\infty} td\mathcal{A}_{0}(t)$, is expressed through
the absorption rate as $\overline{t_{\rm{tr}}} = 1/q$ (here $d\mathcal{A}_{0}
(t)/dt = qe^{-qt}$ is the probability density of the surviving time). At
$t=t_{\rm{tr}}$ an infinite jump of $x(t)$ occurs in the positive or negative
direction, which is interpreted as the transition of a particle to the
absorbing state (see also Ref.~\cite{DKH}). Thus, in this case the sample paths
are the randomly interrupted realizations of the same deterministic function
$z(t)$ $[z(0)=0]$ satisfying the motion equation $\dot{z}(t) = f(z(t),t)$ (in
the case of compound noise the sample paths are the randomly interrupted random
processes, see Sect.~\ref{sec:Comp}). It is this property of the sample paths
which makes it possible to model a number of physical processes interrupted at
random times by the use of the Langevin equation driven by superheavy-tailed
noise.

Examples of these processes can be found, e.g., in systems where the particle
dynamics displays a qualitative change at some random time. One of them, the
particle propagation in an absorbing medium characterized by a constant
absorption rate, is studied in this paper. We note that because all
superheavy-tailed noises generate the absorbing state, for the numerical
simulations of the Langevin equation (\ref{Langevin}) driven by this noise the
probability density $p(y)$ can always be chosen as in (\ref{p2}). The other
example is the position of a fluid particle in a boiling liquid. In this case,
the interruption of the process corresponds to the transition of a particle to
the vapor state. One more example is the position of a free electron in a
semiconductor. Here the interruption of the process occurs due to the
recombination of the electron-hole pair.

\section{Conclusions}
\label{sec:Concl}

We have developed a general approach for studying the statistical properties of
particles driven by superheavy-tailed noises. These noises arise from the noise
generating processes whose independent increments are characterized by the
absence of finite fractional moments. Due to this feature, the distribution of
particles has been studied in detail. Our approach is based on the generalized
Fokker-Planck equation (\ref{FP1}) which corresponds to the Langevin equation
(\ref{Langevin}). Within this framework we have established that the
distribution of particles is the same as in the case of particles moving in an
absorbing medium characterized by the rate of absorption $q$. In other words,
superheavy-tailed noise plays the role of an absorbing medium.

By analyzing the solution of the generalized Fokker-Planck equation, we have
shown that the distribution function can be divided into two parts which
describe two probabilistic states of each particle. These states, surviving and
absorbing, are induced by superheavy-tailed noise and are realized with the
probabilities $e^{-qt}$ and $1-e^{-qt}$, respectively. In the surviving state
the particle dynamics occurs in such a way as if this noise is absent.
Therefore, the distribution of particles in this state is non-degenerate or
degenerate depending on that is the driving noise compound or not. We have
derived the corresponding distribution functions for both these cases. An
infinite jump of a particle, which occurs at a random time under the action of
superheavy-tailed noise, is interpreted as the transition of this particle to
the absorbing state. We have shown explicitly that the distribution of
particles in this state is characterized by the condition that the probability
to find these particles in any finite interval equals zero.

Using the theory of slowly varying functions, we have derived a general
representation for the main characteristic of superheavy-tailed noise, i.e.,
the absorption rate $q$. This representation depends on the asymptotic behavior
of the superheavy-tailed probability density $p(y)$ at $y \to \infty$ and the
scale function $a(\tau)$ at $\tau \to 0$. The asymptotic behavior of $a(\tau)$
is controlled by the asymptotic behavior of $p(y)$, which is assumed to be
known. Finally, as an illustration of our results, we have found the absorption
rate and the scale function for a wide class of superheavy-tailed densities.

\end{document}